\documentclass{pasj00}
\draft

\begin{document}
\SetRunningHead{Y. Shioya et al.}{Intermediate-band Dropout}
\Received{2005/2/5}
\Accepted{2005/2/15}

\title{
The Intermediate-band Dropout Method: A New Method
to Search for High-Redshift Galaxies
}

\author{Yasuhiro \textsc{Shioya} \altaffilmark{1}, 
Yoshiaki \textsc{Taniguchi} \altaffilmark{1}, 
Masaru \textsc{Ajiki} \altaffilmark{1}, 
Tohru \textsc{Nagao} \altaffilmark{1,2}, \\
Takashi \textsc{Murayama} \altaffilmark{1}, 
Shunji \textsc{Sasaki} \altaffilmark{1}, 
Ryoko \textsc{Sumiya} \altaffilmark{1}, \\
and 
Yuichiro \textsc{Hatakeyama}\altaffilmark{1}
}
\email{shioya@astr.tohoku.ac.jp}

\altaffiltext{1}{Astronomical Institute, Graduate School of Science, Tohoku University, \\
                 Aramaki, Aoba, Sendai 980-8578}
\altaffiltext{2}{INAF --- Osservatorio Astrofisico di Arcetri,\\
                 Largo Enrico Fermi 5, 50125 Firenze, Italy}



\maketitle

\begin{abstract}
We propose a new method to search for high-redshift galaxies
that is based on an intermediate-band dropout technique rather than
the usual broad-band dropout one. In this method, we use an 
intermediate-band filter whose central wavelength is longer than
7000 \AA. This new method makes it
possible to distinguish both very late-type stars such as
L and T dwarfs and dusty galaxies at intermediate redshift from real high-$z$ 
Lyman break galaxies.
The reason for this is that such interlopers do not show strong intermediate-band
depression  although they have very red broad-band colors that are indicative
of Lyman break galaxies.
Applying our new method to imaging data sets obtained with the Suprime-Cam
on the Subaru Telescope, we find a new sample of Lyman break galaxies at
$z \simeq 5$. 
\end{abstract}

\section{Introduction}

Recently, much attention has been paid to searches for galaxies
beyond $z=6$ since such galaxies provide us insight on the galaxy formation
(e.g., Hu et al. 2002; Kodaira et al. 2003; Stanway et al. 2004; 
Ajiki et al. 2003, 2004; Taniguchi
et al. 2005; see for a review Taniguchi et al. 2003).
Moreover, there have been reports in the literature that
reionization may occur close to $z=6$ (e.g., Loeb \& Barkana 2001) 
and galaxies rather than QSOs are believed to be the main sources of 
UV photons at those redshifts. 

Based on the color selection
technique,  LBGs at $z \sim 6$ can be selected from deep broad band images
 as galaxies with very
red $i^\prime - z^\prime$ color (e.g., Bouwens et al. 2003; Stanway et al.
 2004; Giavalisco et al. 2004; Dickinson et al. 2004). However, one
problem in this method is that other sources also have such very red colors
in  $i^\prime - z^\prime$; very late-type stars such as L and T dwarfs and
dusty galaxies at intermediate redshifts. With a very high resolution image taken
with the Hubble Space Telescope, one can reject point-source like objects
that should be stars in our Galaxy (Giavalisco et al. 2004).  
However, it is impossible to make such
rejections if we use imaging survey data taken with ground-based telescopes.
Although near infrared data are helpful in selecting real very high-$z$ galaxies,
it seems difficult to carry out wide-field near infrared imaging surveys because
near infrared cameras usually have a small field of view compared 
with those of optical cameras.

In the optical broad-band color selection, only two filters, $i^\prime$, and $z^\prime$,
are used to search for high-$z$ galaxies ($z > 6$). If we can use some other filters at
wavelengths longer than 7000 \AA, we have more useful photometric information
on high-$z$ galaxies, providing us with a much more reliable sample of high-$z$
 galaxy candidates.
Here we note that a set of intermediate-band filters (the IA series)
whose spectral resolution is $R=23$ is available for the Suprime-Cam on the
Subaru Telescope (e.g., Taniguchi 2004; Fujita et al. 2003; Ajiki et al. 2004; 
Yamada et al. 2005; Sumiya et al. 2005; see also Hayashino et al. 2000, 2003).
We here propose to use these 
filters to select high-$z$ LBGs; the intermediate-band dropout method. 
This method uses an imaging data set in which deep intermediate-band
imaging is available together with broad-band data. It is also required that the
band pass of the intermediate-band filter is included in the wavelength coverage of 
one broad band filter.
The merit of this new method is to distinguish both very late-type stars such as
L and T dwarfs and dusty galaxies at intermediate redshift from real high-$z$ LBGs.
The reason for this is that such interlopers do not show strong intermediate-band
depression  although they have very red broad band colors that are indicative of LBGs.

In this Letter, in order to demonstrate that our new method is useful in
finding high-$z$ galaxies, we present our results by using the following two
intermediate-band filters; (1) IA709 ($\lambda_{\rm c} = 708.2$ nm and FWHM $=$ 31.8 nm;
see Yamada et al. 2005; Sumiya et al. 2005), and (2) IA827 ($\lambda_{\rm c} = 827.5$ nm
and  FWHM $=$ 34.0 nm ; see Ajiki et al. 2004). 

\section{The Intermediate-Band Dropout Method}

\subsection{ The IA709 Dropout Method}

First, we describe the IA709 dropout method. 
In this case, we use our deep imaging data taken with the three filters of 
$R$, $i^\prime$, and IA709. In the upper panel of Fig. 1, we show 
the transmission curve of the three filters taken from Miyazaki et al.
(2002)\footnote{See also
http://www.subarutelescope.org/Observing/Instruments/SCam/sensitivity.html}.
In the lower panel of Fig. 1, we also show the spectral energy distribution
(SED) of LBGs at $z = 5.0$ and 5.3 that are generated by using
the population synthesis model, GALAXEV (Bruzual \& Charlot 2003);
see Appendix. 

In Fig. 2, we show the  $R_{\rm C} - i^\prime$ and $IA709 - i^\prime$ diagram.
The left panel shows the loci for model galaxies for the E (Elliptical) model
(red line), SB (starburst) model (the green line), and the SB model with emission lines
(the blue line; see Appendix 1).
We also show the following stellar data points; (1) O to M stars (orange asterisks) that
are calculated by the SEDs given in Gunn \& Stryker (1983), and (2) M9, L0, and L3-L9 dwarfs 
(orange crosses) discussed in Hawley et al. (2002). 
As shown in this panel, if we adopt a criterion of $IA709 - i^\prime >$ 1.5,
we can select LBG candidates at $z \approx 4.9$ to 5.2 as IA709-depressed objects,
being independent of the strength of the Ly$\alpha$ emission. 

In order to demonstrate that this IA709 dropout method works well,
we apply it to the deep imaging data obtained in the MAHOROBA-11 field 
(Yamada et al. 2004; Sumiya et al. 2005). 
Their survey depths (3$\sigma$ limits in a 2$^{\prime \prime}$ diameter aperture) 
are 28.2 ($B$), 27.4 ($R_{\rm C}$), 27.4 ($i^\prime$), 25.8 ($z^\prime$), and 
26.6 ($IA709$). 

It is unlikely that there are galaxies with the elliptical-like SED
at $z > 2$. Therefore, we estimate that a reddest $IA709 - i^\prime$ color 
may be $\approx$ 1.3, corresponding to the color of elliptical galaxies at $z \sim 1.5$. 
Therefore, as noted above, we adopt the following selection criterion for
IA709-dropout galaxies,
\begin{equation}
IA709 - i^\prime > 1.5. 
\end{equation}
In order to secure that our candidates are significantly detected, we also adopt
\begin{equation}
i^\prime < 26.0  \; (7.5 \sigma). 
\end{equation}
In order to reduce any contaminations from foreground objects 
that are free from absorption by the intergalactic neutral hydrogen, 
we also adopt
\begin{equation}
B > 28.2 \; (3 \sigma).
\end{equation}
By using the above three criteria, we first obtain a sample of 35 IA709-dropout objects. 

As shown in Fig. 2, Galactic stars mostly trace the following relation: 
\begin{equation}
(IA709 - i^\prime) \simeq (R - i^\prime) -  0.8, 
\end{equation}
as shown in Fig. 2 (the long dashed line).
Taking account of the scatter of colors of Galactic stars, 
we adopt the following criterion,
\begin{equation}
IA709 - i^\prime > R-i^\prime.
\end{equation}
Then we finally obtain a sample of 19 IA709-dropout galaxies.
These galaxies are shown in the right panel of Fig. 2. 
Their photometric properties are summarized in Table 1. 
In this table, we present $1 \sigma$ error of colors. 
 
For none of these candidates have spectroscopic redshifts been measured till now.
However, since we have multi-bands (4 broad-bands and 7 intermediate-bands) photometric
data of these objects, we can evaluate their photometric redshifts
(the detail of our method is written in Appendix 2).
We then find that 18 among the 19 IA709-dropout galaxies have $z \approx$ 4.9 to 5.2.
One object is identified as an E galaxy at $z \approx 1.5$ although its real redshift
is unknown. 
We therefore conclude that the IA709-dropout method is useful 
in searching for high-redshift ($z \simeq 5$) galaxies within a 
relatively small redshift range. 

It is worthwhile studying whether or not we can select the above IA709 dropouts
by using only broad-band photometric data.
In the right panel of Fig.3, we plot 19 IA709-dropout galaxies on the $R_{\rm C}-i^\prime$
vs. $IA709-i^\prime$ diagram. The left panel shows only the model loci together with 
the stellar data given in Fig. 2. As shown in this figure, it seems 
difficult to select our sample galaxies at $4.9 < z < 5.2$ solely from this diagram. 

\subsection{The IA827 Dropout Method}

Next, we describe the IA827-dropout method. 
This idea was first discussed by Fujita (2003). 
In this case, we use our deep imaging data taken with filters of 
$I_{\rm C}$, $z^\prime$, and IA827 taken with Ajiki et al. (2003, 2004). 
In the upper panel of Fig. 4, we show the transmission curve of 
the three filters taken from Miyazaki et al. (2002).
In the lower panel of Fig. 4, we show SEDs of both a model LBG
at $z = 6.1$, dusty starburst with $A_V=10$ mag at $z = 1.3$ 
and the Galactic T dwarf star 2MASS J04151954-0935066 
(Kirkpatrick 2003). 
We use the starburst reddening curve of Calzetti et al. (2000) 
to calculate the SED of a dusty starburst galaxy. 
It is shown that LBGs at $z \sim 6$ can be selected as an IA827 dropout.
However, neither dusty starbursts nor T dwarfs can be selected as an IA827 
dropout even though they have a very red  $I_{\rm C} - z^\prime$  color. 
In this way, we can select LBGs at $z>5.8$ by using the IA827-dropout method.

In order to demonstrate that this IA827 dropout method works well,
we apply it to the optical imaging data of SDSS J104433.04$-$012522.2
(Ajiki et al. 2003, 2004).
Their survey depths under a 2$^{\prime\prime} ~ \phi$ aperture and 3$\sigma$ conditions)
are 26.6 ($B$), 26.2 ($R_{\rm C}$), 25.9 ($I_{\rm C}$), 25.3 ($z^\prime$), and 
25.6 ($IA827$). 
In Fig. 5, we show the diagram between  $I_{\rm C} - z^\prime$ and $IA827 - z^\prime$
for galaxies found in Ajiki et al. (2004). 
We show loci of the three types of model galaxies that are generated by
the same method given in Fig. 2.
We also show the data points for Galactic stars (Gunn \& Stryker 1983) and 
L and T dwarfs (Kirkpatrick 2003)\footnote{
http://spider.ipac.caltech.edu/staff/davy/ARCHIVE/}.
 
As shown in Fig.5, being independent of the strength of the Ly 
$\alpha$ emission, galaxies at $5.9 < z < 6.2$ can be identified as 
IA827-dropout objects. 
Assuming again that there are no elliptical-like red galaxies at $z \sim 2$, 
we estimate a reddest $IA827 - z^\prime$ color of $\approx$ 1.4,
corresponding to the color of elliptical galaxies at $z \sim 2$. 
Therefore, we adopt the following selection criterion for
IA827-dropout galaxies,
\begin{equation}
IA827 - z^\prime > 1.5
\end{equation}
together with
\begin{equation}
z^\prime < 25.3 \; (3 \sigma). 
\end{equation}

In order to reduce any contaminations from foreground objects 
that are free from absorption by the intergalactic neutral hydrogen, 
we also adopt following criteria
\begin{eqnarray}
B & > & 26.6 \; (3 \sigma) \\
R_{\rm C} & > & 26.2 \; (3 \sigma)
\end{eqnarray}
By using the above three criteria, we first obtain a sample of 9 IA827-dropout objects. 

Next, we must separate high-$z$ galaxies from Galactic stars including L and T dwarfs.
Since such stars trace the following relation,

\begin{equation}
(IA827 - z^\prime) \simeq (I_{\rm C} - z^\prime) - 0.5, 
\end{equation}
we adopt the following criterion to separate LBG candidates from them,
\begin{equation}
IA827 - z^\prime > I_{\rm C} - z^\prime,
\end{equation}
where the scatter of colors of stars is taken into account again.
Then we finally obtain a sample of 6 IA827-dropout galaxies. 
These galaxies are shown in Fig. 5 by open boxes. 
Their basic data are summarized in Table 2. 

It is noted that four among the six IA827-dropout candidates are brighter than $z^\prime = 25$.
They may be too bright for such very high-$z$ galaxies. 
This data set is shallower in the survey depth than that of MAHOROBA-11 
and considered to be too shallow to select convincing high-z galaxies on this occasion. 
Therefore follow-up optical
spectroscopy will be necessary to confirm their redshifts.

\section{REMARKS}

We have demonstrated that the IA709-dropout method is useful in
finding LBGs at $4.9 < z < 5.2$. Since this redshift range is 
relatively small, this method will be useful to isolate such LBGs
in some targeted sky areas.
On one hand, the IA827-dropout method is useful in finding LBGs at $z > 6$.
Although we often adopt a color criterion of $i^\prime - z^\prime >$ 1.3 -- 1.5 
to search for LBGs at $z > 6$ (e.g., Nagao et al. 2004; Stanway et al. 2004;
Giavalisco et al. 2004), this criterion is not enough to select such LBGs
unambiguously because L and T dwarfs also have such very red colors.
Therefore, the most important merit of the IA827-dropout method is 
that major interlopers, L and T dwarfs, are unambiguously separated from real LBGs.
The IA709 dropout method also has this method.
Hu et al. (2004) noted that LAEs at $z \simeq 5.7$ are well separated from 
L and T dwarfs if a proper criterion on the $NB816 - z^\prime$ color is adopted. 
We also note that Kakazu et al. (2004) proposed the NB816-depressed method to 
separate high-redshift quasars and galaxies from L and T dwarfs based on 
the same principle. 

We have shown that the use of intermediate-band filters gives a good advantage
over using only broad-band data. The broad-band color selection method is
useful in finding a larger number of high-$z$ galaxies because the survey volume
is wider. However, this method suffers from contamination of objects whose
broad band colors are similar to those of high-$z$ galaxies; e.g., L and T dwarfs,
and dusty starburst galaxies at intermediate redshift.
On one hand, the Ly$\alpha$ selection method by using a narrowband filter is 
basically free from such contamination. Further, this method enables us to
find fainter high-$z$ galaxies than the broad-band color selection method
(e.g., Fynbo et al. 2003). However, follow-up spectroscopy is
necessary to exclude strong emitters at lower redshifts. It is also noted that
the survey volume is much smaller than that of the broad-band color selection.
The use of an intermediate-band filter at $\lambda > 7000$ \AA ~ together with
broad band ones makes it possible to exclude contaminations from L and T dwarfs.
The survey volume is wider than that based on the narrowband selection.
Strong Ly$\alpha$ emitters can be also found in a survey with an intermediate-band
filter (e.g., Fujita et al. 2003; Ajiki et al. 2004; Yamada et al. 2005; Sumiya
et al. 2005).


We would like to thank the staff of the Subaru Telescope. 
We would also like to thank the referee, Johan Fynbo, for very careful and thoughtful report.
This work was financially supported in part by
the Ministry of Education, Culture, Sports, Science, and Technology
(Nos. 10044052, and 10304013) and JSPS (No. 15340059).
MA and TN are JSPS fellows.

\newpage

\appendix
\section{The SED model of galaxies}

We use the population synthesis model, GALAXEV (Bruzual \& Charlot 2003) 
to generate model galaxy SEDs. 
The SEDs of local galaxies are well reproduced 
by models whose star-formation rate declines exponentially 
(the $\tau$ model): i.e., $SFR(t) \propto {\rm exp}(-t/\tau)$, where $t$ is the
age of galaxy and $\tau$ is the time scale of star formation.
In this study, we use $\tau=1$ Gyr models with Salpeter's initial mass function
(the power index of $x=1.35$ and the stellar mass range of 
$0.1 \leq m/M_{\odot} \leq 100$) to derive various SED types.
We note that the SED templates derived by Coleman et al. (1980) for  
elliptical galaxies, Sbc, Scd, Irr, and SB, correspond to those using 
$t=$8, 4, 3, 2, and 1 Gyr, respectively. 
In this paper, we therefore called SEDs with age of $t=$8, 4, 3, 2, and 1 Gyr 
Ell type, Sbc type, Scd type, Irr type, and SB type, respectively. 
Some of high-redshift galaxies show strong Ly$\alpha$ emission line, 
which may modify the observed colors of galaxies. 
In order to include possible emission-line fluxes into our models,
we calculate the number of ionizing photons, $N_{\rm Lyc}$, 
from the SED which is calculated from the above population synthesis model. 
Then we can estimate H$\beta$ luminosity, $L$(H$\beta$),
using the following formula (Leitherer \& Heckman 1995), 

\begin{equation}
L(H\beta) = 4.76 \times 10^{-13} N_{\rm Lyc}\hspace{3mm} \rm erg \hspace{1mm} s^{-1}.
\end{equation}
Other strong emission-line luminosities, such as for [O {\sc ii}], 
[O {\sc iii}], H$\alpha$, and so on, are estimated by typical line ratios
relative to H$\beta$ (PEGASE: Fioc \& Rocca-Volmerange 1997). 

\section{Photometric Redshifts of IA709 dropout objects}

A photometric redshift of a galaxy was evaluated from the global shape of 
the SED of a galaxy, as a redshift with the maximum likelihood, 
\begin{equation}
L(z,T) = \prod_{i=1}^{11} \exp \{ -\frac{1}{2} \left[ \frac{f_i - A F_i(z,T)}{\sigma_i} \right]^2 \}
\end{equation}
where , $f_i$, $F_i(z,T)$ and $\sigma$ are the observed flux, 
the template flux of SED type $T$ and redshift $z$, and the error of the $i$-th band, 
respectively, and $A$ is defined as 
\begin{equation}
A = \frac{\sum F_if_i/\sigma_i^2}{\sum F_i^2/\sigma_i^2}.
\end{equation}
The SED of a galaxy that we observed was mainly determined by the following four factors: 
(1) the radiation from stars and ionized gas in the galaxy, (2) the extinction by dust in the galaxy, itself, 
(3) the redshift of the galaxy, and (4) the absorption by the intergalactic neutral hydrogen 
between the galaxy and us. We treated the above factors in the following way. 

The SED model of galaxies is written in Appendix 1. 
We adopted the dust extinction curve for starburst galaxies determined by Calzetti et al. 
(2000) with visual extinctions of 0.0, 0.1, 0.3, 0.5, 1.0, and 2.0. 
As for the absorption by intergalactic neutral hydrogen, we used the average optical 
depth derived by Madau et al. (1996). 
We then estimated the probable photometric redshifts of the galaxies. 
In this procedure, we adopted an allowed redshift of between $z=0$ and $z=7$ 
with a redshift bin of $\Delta z = 0.01$.


\clearpage

\begin{table}
\begin{center}
\caption{Photometric properties of IA709 dropout galaxy candidates}
\begin{tabular}{ccccccccc}
\hline
\hline
  ID    & \multicolumn{4}{c}{Optical AB magnitude} &  \multicolumn{3}{c}{Optical colors}                     & $z_{\rm phot}$\\
        & $R_C$ & $i^\prime$ & $z^\prime$ & IA709 & $R_C-i^\prime$ & $IA709-i^\prime$ & $i^\prime - z^\prime$&               \\
\hline
   32769& 27.08& 25.70& 25.00& 27.23& $1.38_{-0.26}^{+0.34}$& $1.53_{ -0.51}^{+ 1.00} $& $ 0.69_{-0.19}^{+0.23}$& 1.49\\
   54476& 26.77& 25.51& 24.86& 27.04& $1.27_{-0.20}^{+0.25}$& $1.53_{ -0.45}^{+ 0.77} $& $ 0.65_{-0.17}^{+0.19}$& 5.16\\
   56089& 27.09& 25.98& 25.99& 27.83& $1.11_{-0.27}^{+0.36}$& $1.85_{ -0.78}^{+\infty}$& $-0.02_{-0.38}^{+0.59}$& 4.98\\
   57515& 26.51& 24.75& 24.11& 26.60& $1.76_{-0.15}^{+0.18}$& $1.85_{ -0.31}^{+ 0.44 }$& $ 0.65_{-0.09}^{+0.09}$& 5.10\\
   61765& 27.21& 25.76& 26.02& 27.75& $1.46_{-0.29}^{+0.39}$& $1.99_{ -0.73}^{+ 3.66 }$& $-0.26_{-0.38}^{+0.59}$& 4.98\\
   62363& 27.17& 25.73& 25.64& 27.62& $1.45_{-0.28}^{+0.37}$& $1.90_{ -0.67}^{+ 2.14 }$& $ 0.09_{-0.29}^{+0.40}$& 4.91\\
   62385& 27.58& 25.41& 24.82& 29.21& $2.17_{-0.37}^{+0.55}$& $3.80_{ -1.68}^{+\infty}$& $ 0.59_{-0.16}^{+0.18}$& 5.01\\
   63164& 27.30& 25.99& 25.52& 28.48& $1.31_{-0.31}^{+0.44}$& $2.49_{ -1.15}^{+\infty}$& $ 0.47_{-0.28}^{+0.37}$& 4.92\\
   67118& 28.14& 25.85& 25.60&(27.8)& $2.29_{-0.56}^{+1.21}$& $>1.95                  $& $ 0.25_{-0.29}^{+0.39}$& 5.07\\
   67913& 27.80& 25.99& 25.81& 28.68& $1.81_{-0.44}^{+0.75}$& $2.69_{ -1.29}^{+\infty}$& $ 0.18_{-0.34}^{+0.49}$& 5.08\\
   67992& 27.31& 25.39& 25.87& 28.26& $1.91_{-0.30}^{+0.41}$& $2.87_{ -1.01}^{+\infty}$& $-0.48_{-0.34}^{+0.49}$& 5.12\\
   71545& 26.57& 25.39& 24.89& 27.00& $1.18_{-0.17}^{+0.21}$& $1.61_{ -0.43}^{+ 0.72 }$& $ 0.50_{-0.16}^{+0.19}$& 5.14\\
   75734& 26.90& 25.71& 25.04& 28.56& $1.19_{-0.29}^{+0.29}$& $2.85_{ -1.21}^{+\infty}$& $ 0.68_{-0.19}^{+0.23}$& 5.06\\
   77639& 27.25& 25.86& 25.17& 27.86& $1.39_{-0.30}^{+0.41}$& $2.00_{ -0.79}^{+\infty}$& $ 0.69_{-0.22}^{+0.27}$& 4.97\\
   78331& 27.23& 25.92& 25.18& 28.30& $1.31_{-0.29}^{+0.40}$& $2.39_{ -1.04}^{+\infty}$& $ 0.73_{-0.22}^{+0.28}$& 5.27\\
   80723& 26.68& 25.32& 25.08& 27.18& $1.36_{-0.19}^{+0.22}$& $1.86_{ -0.49}^{+ 0.92 }$& $ 0.25_{-0.18}^{+0.22}$& 5.21\\
   81057& 27.36& 25.71& 25.14& 28.00& $1.65_{-0.32}^{+0.45}$& $2.29_{ -0.86}^{+\infty}$& $ 0.57_{-0.21}^{+0.25}$& 5.12\\
   85959& 26.77& 25.47& 25.27& 27.04& $1.29_{-0.20}^{+0.25}$& $1.57_{ -0.45}^{+ 0.77 }$& $ 0.20_{-0.22}^{+0.27}$& 5.01\\
   91190& 27.21& 25.85& 25.70& 28.88& $1.36_{-0.29}^{+0.39}$& $3.03_{ -1.42}^{+\infty}$& $ 0.15_{-0.30}^{+0.42}$& 4.93\\
\hline
\end{tabular}
\end{center}
\end{table}

\begin{table}
\begin{center}
\caption{Photometric properties of IA827-dropout galaxy candidates}
\begin{tabular}{cccccc}
\hline
\hline
  ID    & \multicolumn{3}{c}{Optical AB magnitude} &  \multicolumn{2}{c}{Optical colors} \\
        & $I_C$ & $IA827$ & $z^\prime$ & $IA827-z^\prime$ & $i^\prime - z^\prime$ \\
\hline
    5873&   26.15 &  26.90 &  25.10  &  $1.80_{-0.81}^{+\infty}$  &  $1.05_{-0.45}^{ +0.79 }$\\
    9143&   25.82 &  26.55 &  24.98  &  $1.57_{-0.65}^{ +1.86 }$  &  $0.84_{-0.37}^{ +0.57 }$\\
   11293&   27.88 &  28.93 &  25.17  &  $3.76_{-2.25}^{+\infty}$  &  $2.71_{-1.24}^{+\infty}$\\
   18175&   25.93 &  26.52 &  24.90  &  $1.62_{-0.64}^{ +1.72 }$  &  $1.03_{-0.38}^{ +0.60 }$\\
   18313&   28.43 &  28.81 &  24.58  &  $4.23_{-2.15}^{+\infty}$  &  $3.85_{-1.64}^{+\infty}$\\
   21239&   25.77 &  25.85 &  24.32  &  $1.53_{-0.39}^{ +0.61 }$  &  $1.45_{-0.31}^{ +0.44 }$\\
\hline
\end{tabular}
\end{center}
\end{table}

\clearpage

\begin{figure}
\begin{center}
\FigureFile(80mm,80mm){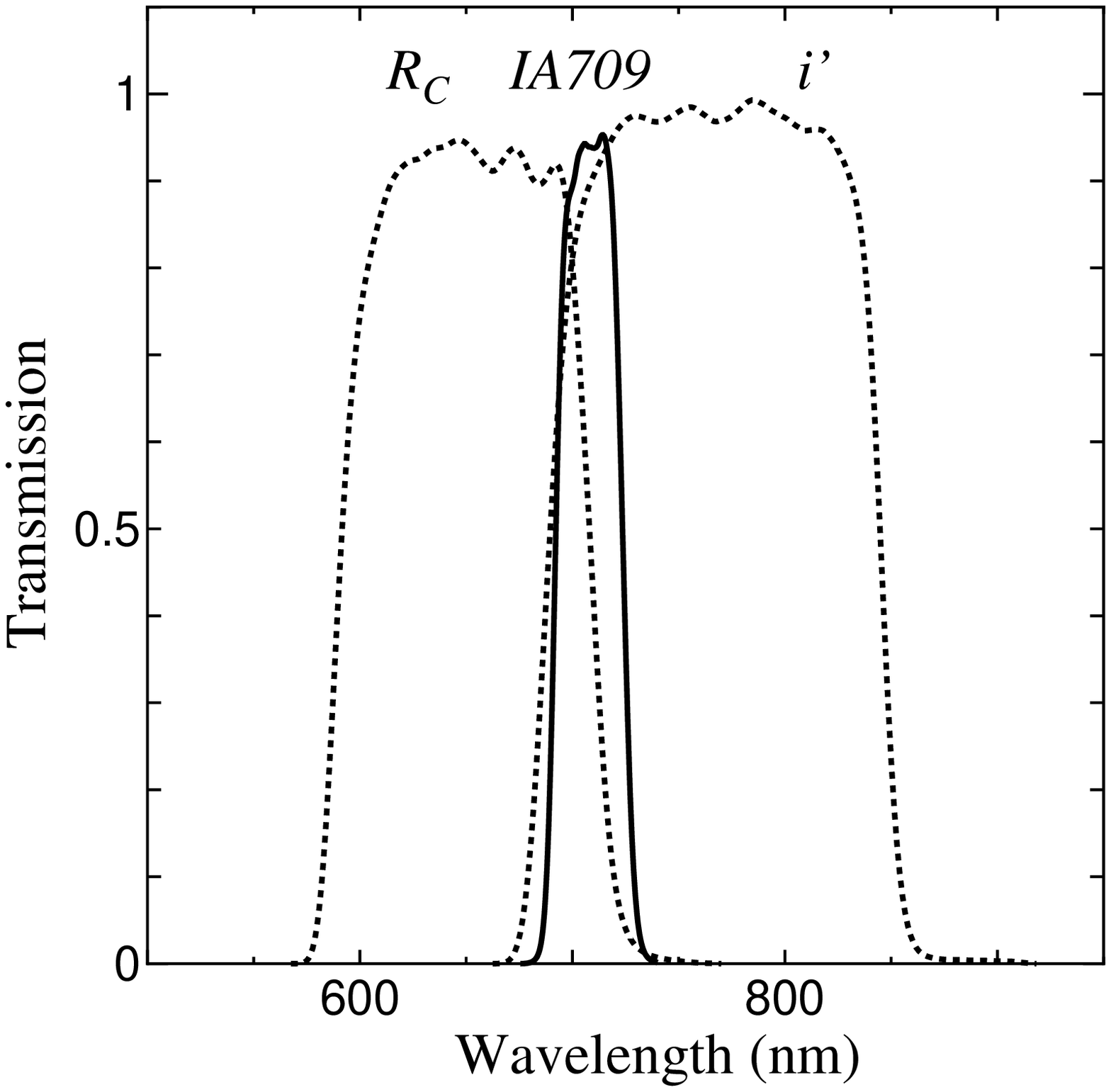}

\FigureFile(80mm,80mm){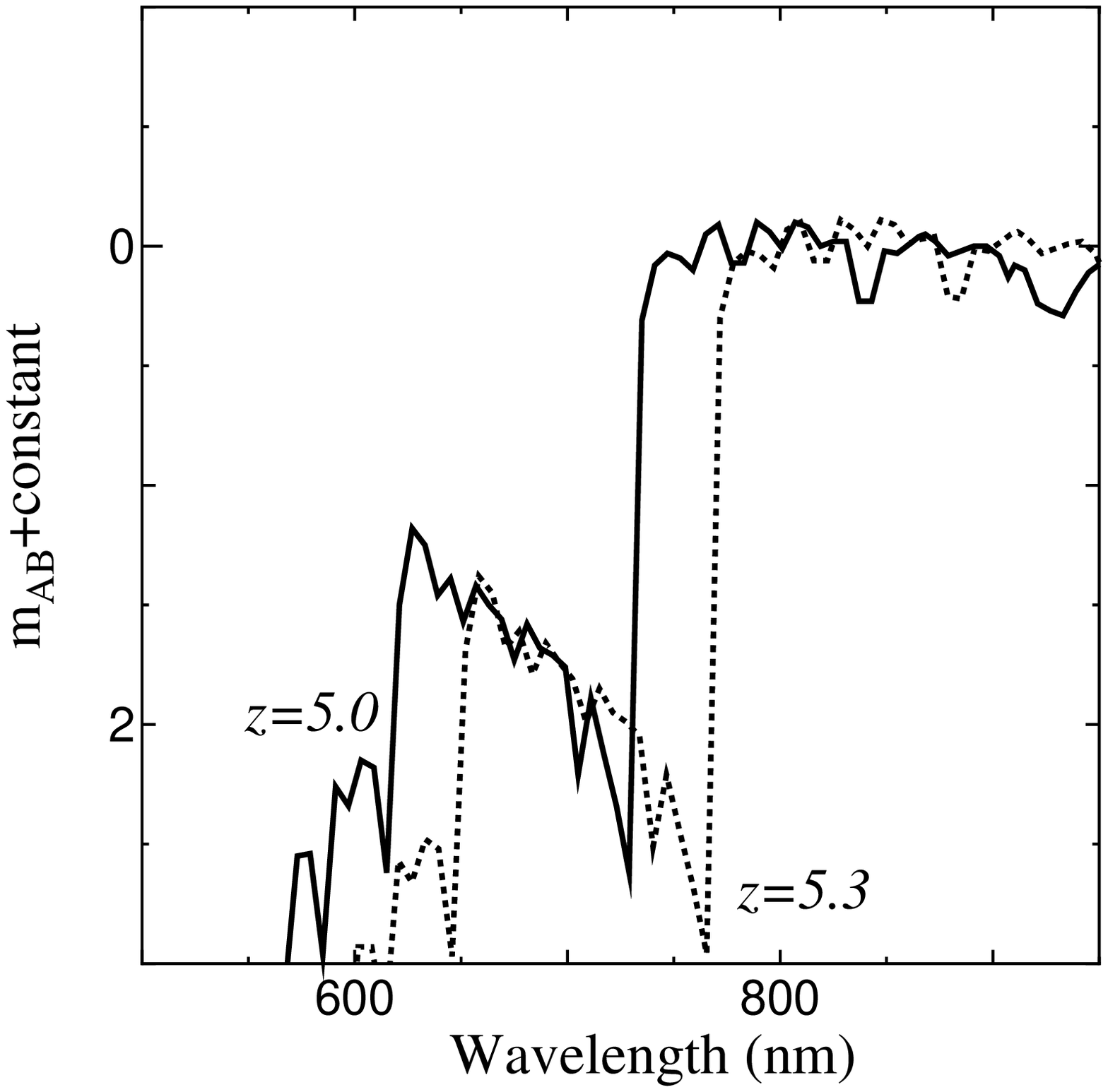}
\end{center}
\caption{
The upper panel shows the transmission curves of the filters, 
$R_C$, $i^\prime$, and IA709. 
The lower panel shows the SED of 
Lyman break galaxies at $z \sim 5.0$ and 5.3.
}\label{fig:fig1}
\end{figure}

\begin{figure}
\begin{center}
\FigureFile(80mm,80mm){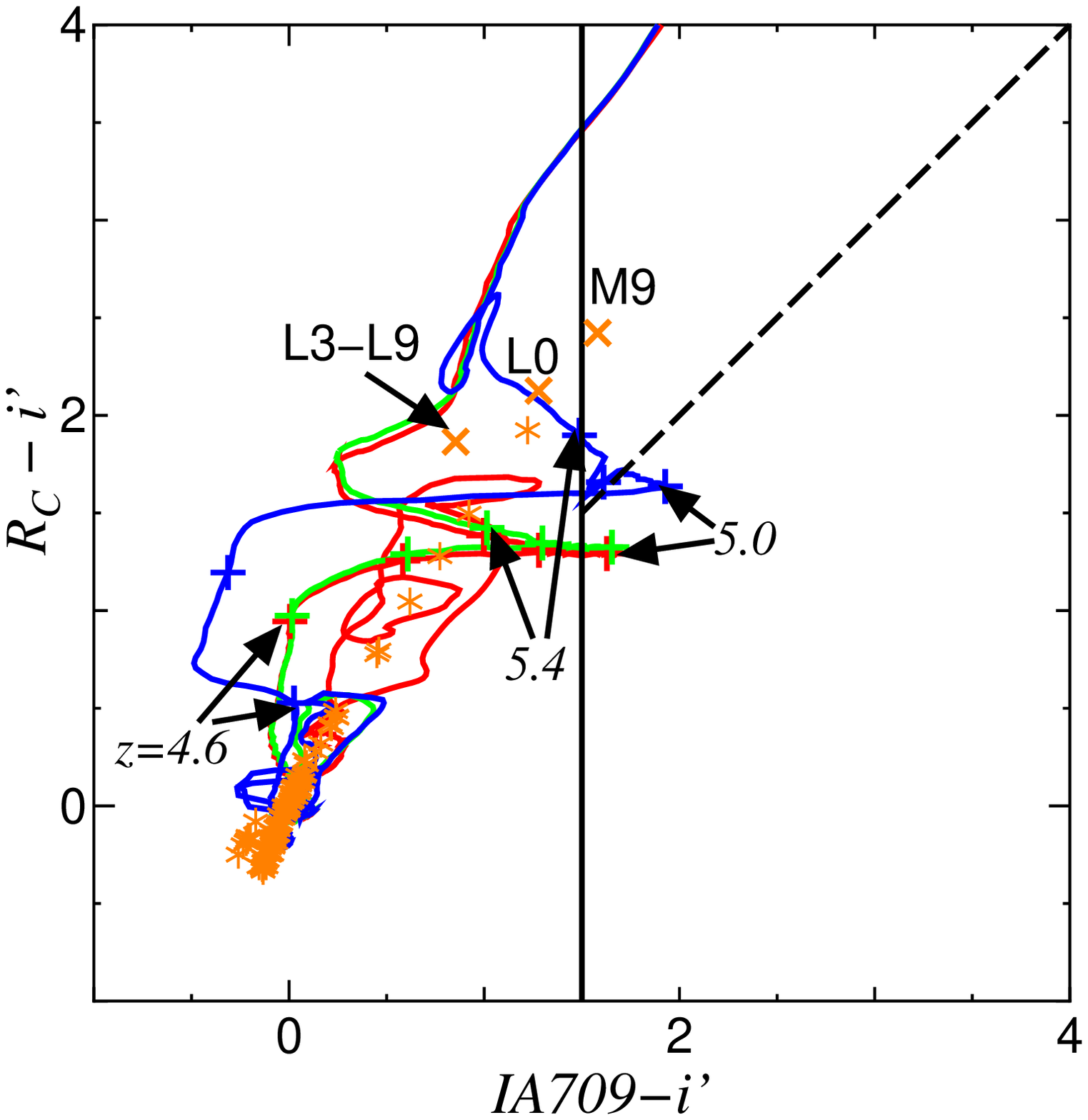}
\FigureFile(80mm,80mm){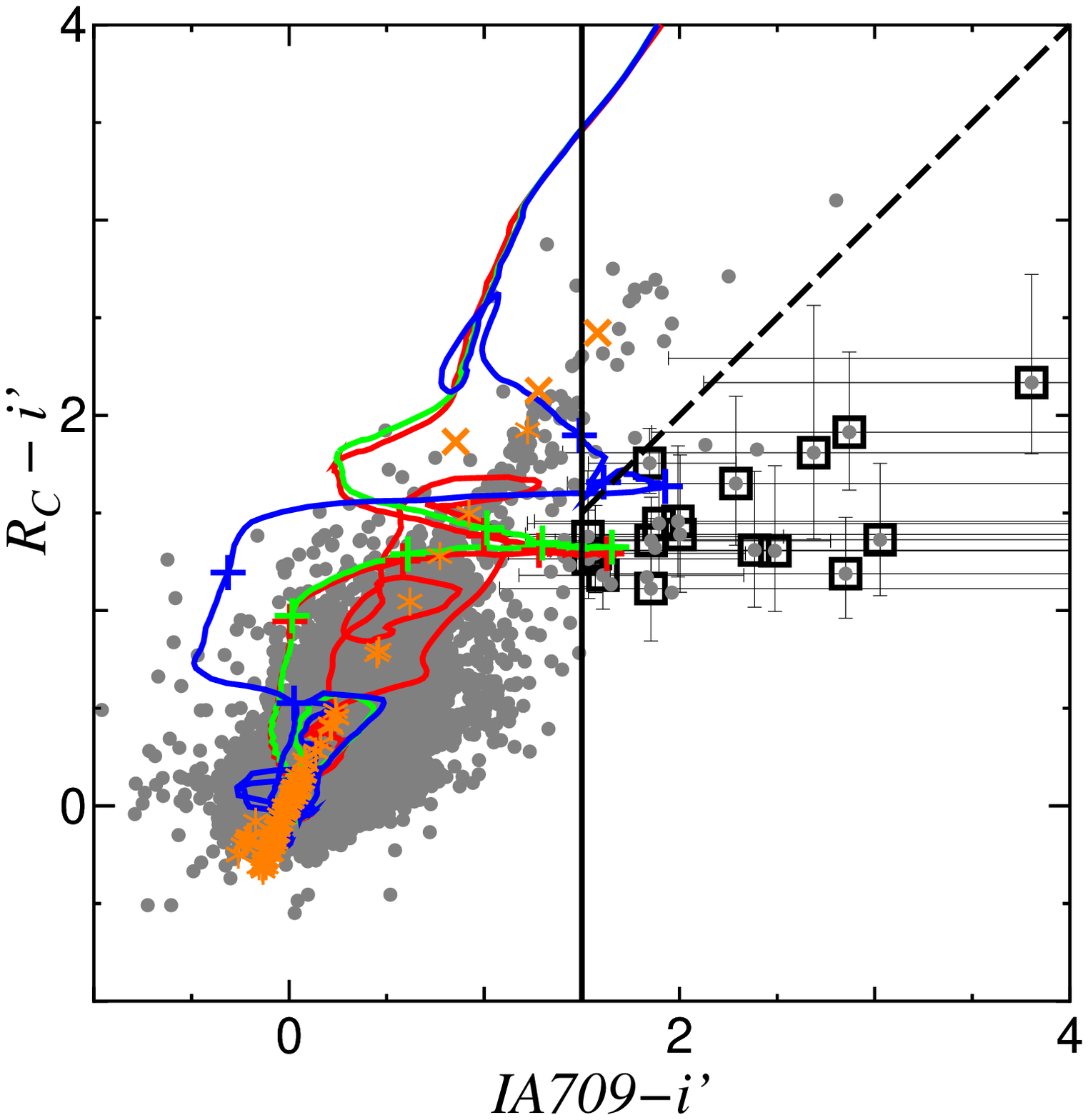}
\end{center}
\caption{
$R_C - i^\prime$ vs. $IA709 - i^\prime$ diagram. 
{\it Left panel}: 
Solid lines are loci of model galaxies produced by GALAXEV (Bruzual \& Charlot 2003): 
red line - $\tau = 1$ Gyr model with age of 8 Gyr, 
green line - $\tau = 1$ Gyr model with age of 1 Gyr (SB model), 
and blue line - SB model with emission lines (see text). 
Stellar data are also shown; orange asterisks are from Gunn \& Stryker (1983)
and orange crosses are from Hawley et al. (2002); see text for details.
{\it Right panel}: 
Our observational data are overlaid on the diagram. 
Gray filled circles show all objects with $20 < i^\prime < 26$. 
Open squares with error bars show our final sample.
}\label{fig:fig2}
\end{figure}

\begin{figure}
\begin{center}
\FigureFile(80mm,80mm){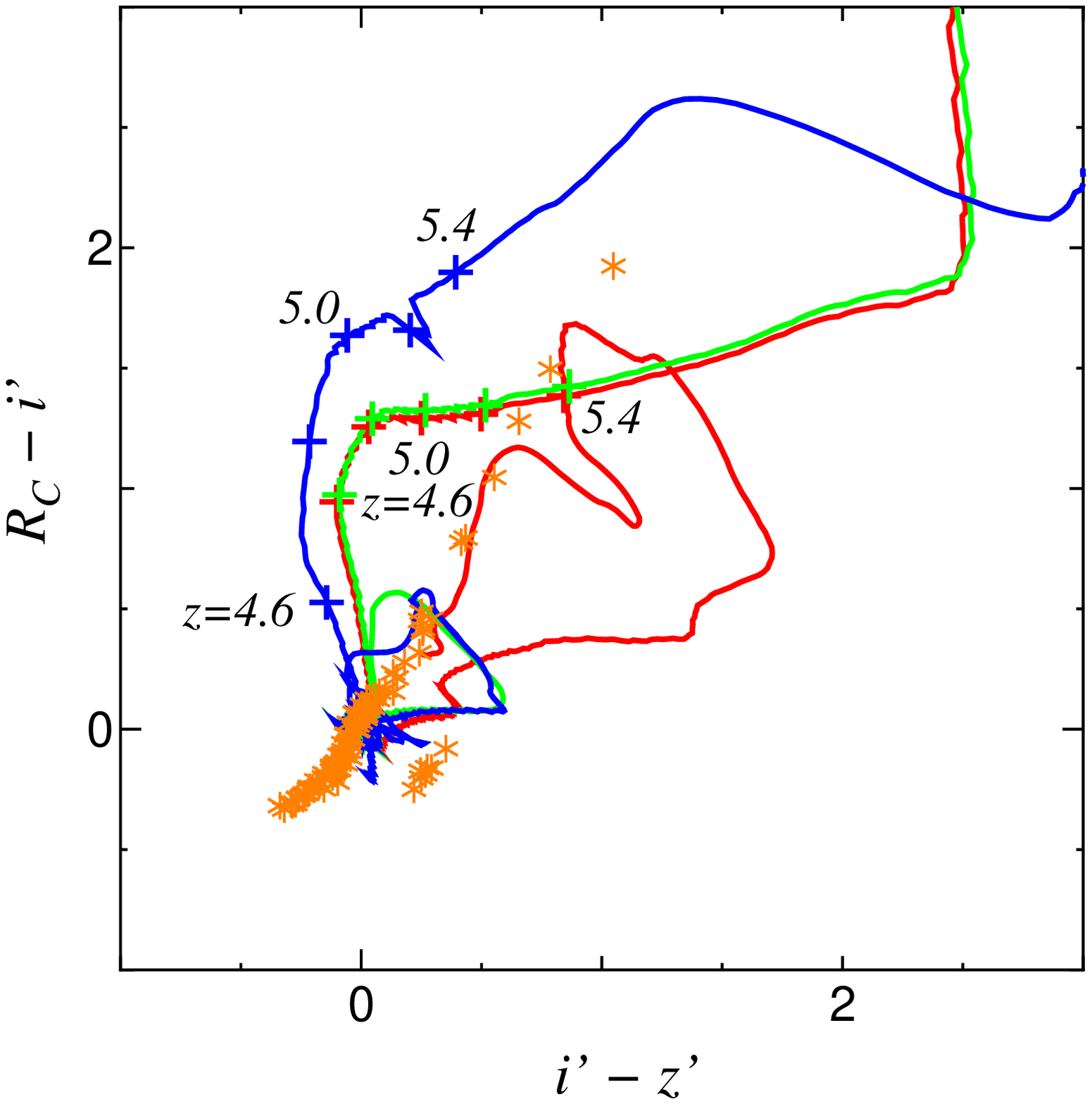}
\FigureFile(80mm,80mm){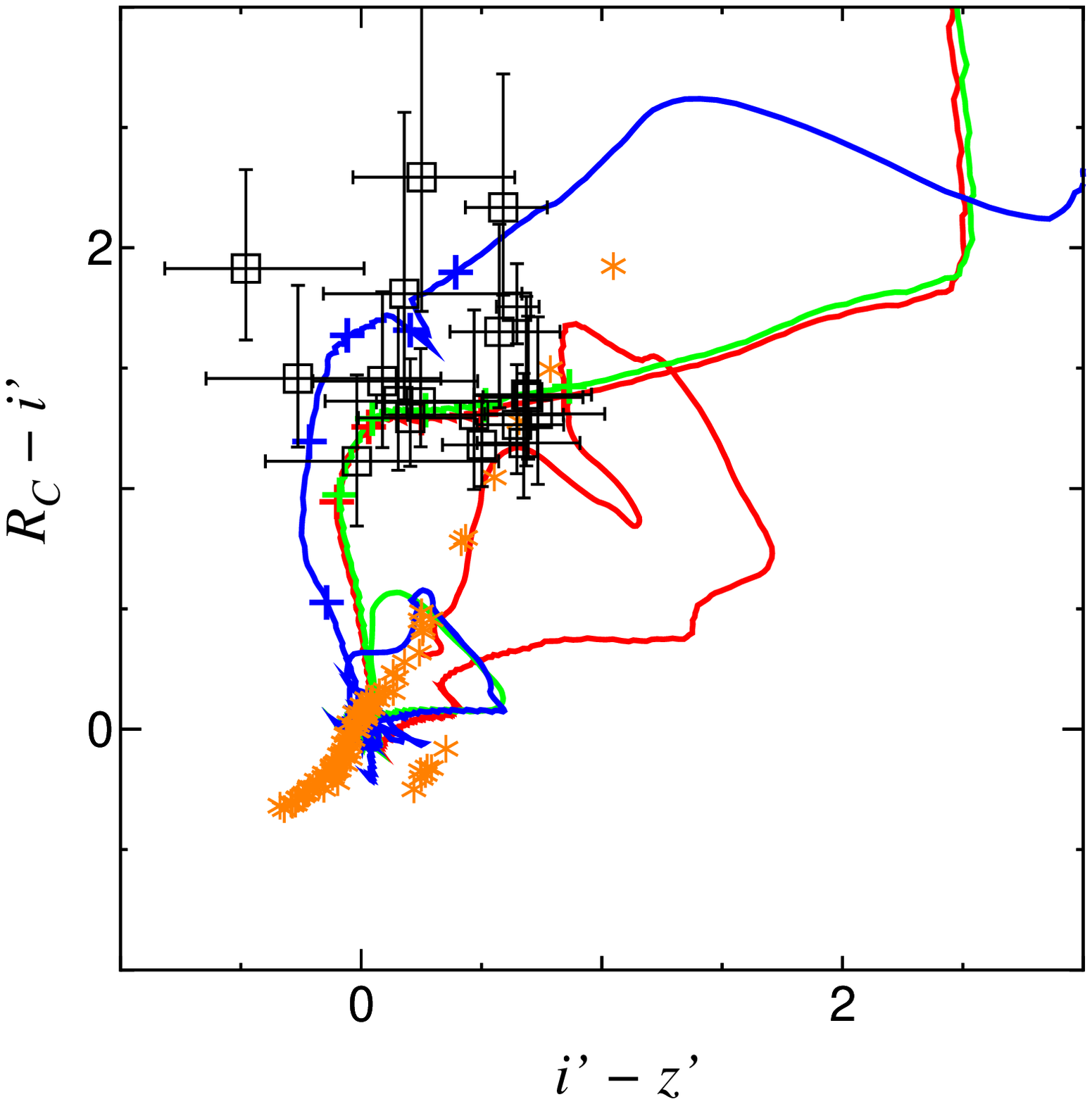}
\end{center}
\caption{
$R_C - i^\prime$ vs. $i^\prime - z^\prime$ diagram. 
{\it Left panel}: The meaning of red, blue and green lines and orange asterisks is 
the same as these given in Fig.2. 
{\it Right panel}: Open squares with error bars show our final sample.
}\label{fig:fig3}
\end{figure}

\begin{figure}
\begin{center}
\FigureFile(80mm,80mm){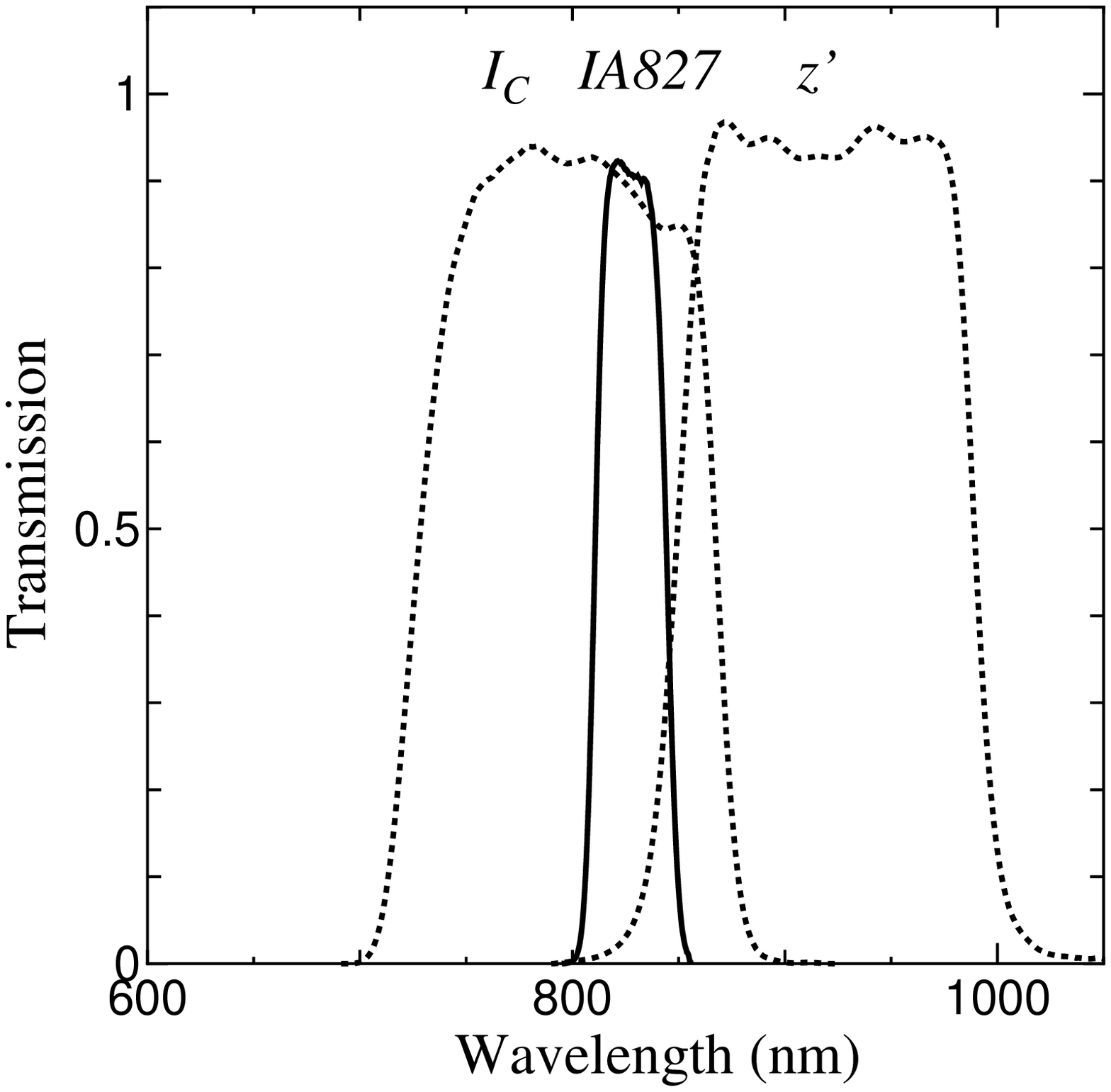}

\FigureFile(80mm,80mm){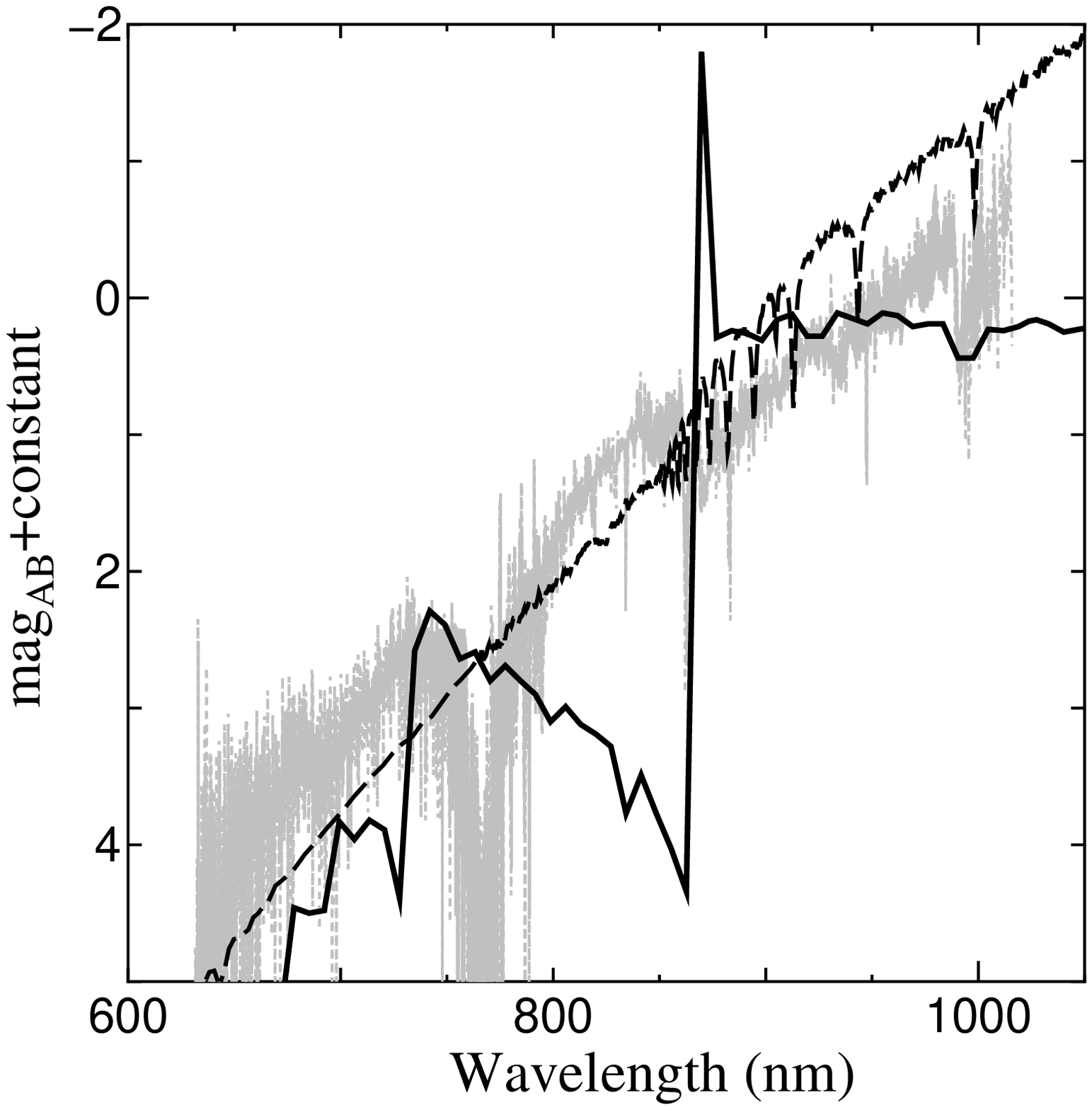}
\end{center}
\caption{
The upper panel shows the transmission curves of the filters: 
$I_C$, $z^\prime$ and IA827. 
The lower panel shows the SED of 
a Lyman break galaxy (LBG) at $z \sim 6.1$ (black solid line) 
dusty starburst with $A_V=10$ mag at $z \sim 1.3$ (black dashed line), 
and the Galactic T dwarf star, 2MASS J04151954-0935066 (gray dotted line; Kirkpatrick 2003). 
The SEDs of an LBG 
and a dusty starburst galaxy are 
produced by GALAXEV 
(Bruzual \& Charlot 2003, see text). 
Although, the two objects shows the same $I - z^\prime$ colors ($\simeq 2$), 
the $IA827-z^\prime$ color of LBG is redder than that of 
an L dwarf star 
and a dusty starburst galaxy.
}\label{fig:fig4}
\end{figure}

\begin{figure}
\begin{center}
\FigureFile(80mm,80mm){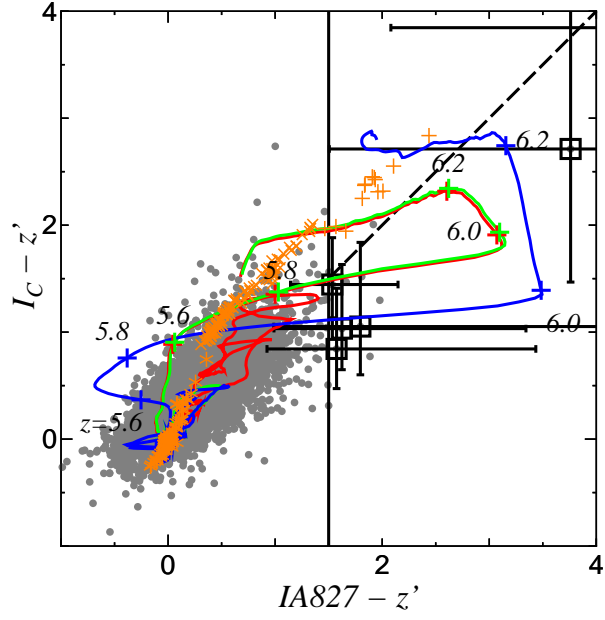}
\end{center}
\caption{
$I_C - z^\prime$ vs. $IA827-z^\prime$ diagram. 
Solid lines are loci of model galaxies produced by GALAXEV (Bruzual \& Charlot 2003): 
red line - $\tau = 1$ Gyr model with age of 8 Gyr, 
green line - $\tau = 1$ Gyr model with age of 1 Gyr (SB model), 
and blue line - SB model with emission lines (see text). 
Orange marks show stars. Asterisks are Gunn \& Stryker (1983)'s stars, 
crosses are L dwarfs and pluses are T dwarfs.
Open squares with error bars show our final sample.
}\label{fig:fig5}
\end{figure}

\end{document}